\documentclass[12pt,a4paper]{article}
\usepackage{amssymb,amsmath,stmaryrd}
\usepackage{epsf}

\newcounter{example}

\newtheorem{define}{Definition}

\newtheorem{prop}{Proposition}

\newtheorem{cor}{Corollary}
\newtheorem{lemma}{Lemma}

\newcommand{\pa}[1] { \left\{ {#1} \right\}  }
\newcommand{\pp}[1] { \left( {#1} \right)  }
\newcommand{\Sipos}{\v{S}ipo\v{s}}
\def \Sint {\check{\mathcal{S}}}
\def \bone   {\boldsymbol{1}}
\def \bzero   {\boldsymbol{0}}
\newcommand{\svee}{\operatornamewithlimits{\varovee}}

\newcommand{\bdot}{\operatornamewithlimits{\boxdot}}

\newcommand{\idfont}[2] {
                        \ifnum #2 = 8  \scriptsize   \fi
                        \ifnum #2 = 10 \small        \fi
                        \ifnum #2 = 11 \footnotesize \fi
                        \ifnum #2 = 12 \normalsize   \fi 
                        \ifnum #2 = 14 \large        \fi
                        \ifnum #2 = 17 \Large        \fi
                        \ifnum #2 = 18 \LARGE        \fi
                        \ifnum #2 = 20 \Huge         \fi}  
\begin{document}

\title{On the Extension of Pseudo-Boolean Functions for the
Aggregation of Interacting Criteria}

\author{Michel GRABISCH\thanks{Corresponding author. On leave from Thomson-CSF, Corporate Research
  Lab, 91404 Orsay Cedex, France}\\
  LIP6\\
  University of Paris VI\\
  4, Place Jussieu, 75252 Paris, France\\
  \normalsize email \texttt{Michel.Grabisch@lip6.fr}
  \and
  Christophe LABREUCHE\\
Thomson-CSF, Corporate Research Laboratory  \\
Domaine de Corbeville, 91404 Orsay Cedex, France \\
\normalsize email \texttt{\{labreuche\}@lcr.thomson-csf.com}
\and
Jean-Claude VANSNICK\\
University of Mons-Hainaut\\
Place du Parc, 20, B-7000 Mons, Belgium\\
\normalsize email \texttt{Jean-Claude.Vansnick@umh.ac.be}}

\date{}

\maketitle

\begin{abstract}
The paper presents an analysis on the use of integrals defined for non-additive
measures (or capacities) as the Choquet and the \Sipos{}
integral, and the multilinear model, all seen as extensions of pseudo-Boolean
functions, and used as a means to model interaction between criteria in a
multicriteria decision making problem. The emphasis is put on the use, besides
classical comparative information, of information about difference of
attractiveness between acts, and on the existence, for each point of
view, of a ``neutral level'', allowing to introduce the absolute notion of
attractive or repulsive act.  It is shown that in this case, the \Sipos{}
integral is a suitable solution, although not unique. Properties of the
\Sipos{} integral as a new way of aggregating criteria are shown, with emphasis
on the interaction among criteria.
\end{abstract}
 

\section{Introduction}
Let us consider a decision making problem, of which the structuring
phase has led to the identification of a family
$\mathcal{C}=\{C_1,\ldots,C_n\}$ of $n$ fundamental \emph{points of view}
(\emph{criteria}), which permits to meet the concerns of the decision maker
(DM) in charge of the above mentioned (decision making) problem. We
suppose hereafter that, during the structuring phase, one has
associated to each point of view $C_i$, $i=1,\ldots,n$, a \emph{descriptor}
(\emph{attribute}), that is, a set $X_i$ of reference levels intended to
serve as a basis to describe plausible impacts of potential actions
with respect to $C_i$. 

We make also the assumption that, for all $i=1,\ldots,n$, there exists in
$X_i$ two particular elements which we call ``$\text{Neutral}_i$'' and 
``$\text{Good}_i$'', and denoted $\bzero_i$ and $\bone_i$ respectively,
which have an absolute signification: $\bzero_i$ is an element which is
thought by the DM to be neither good nor bad, neither attractive nor
repulsive, relatively to his concerns with respect to $C_i$, and
$\bone_i$ is an element which the DM considers as good and completely
satisfying if he could obtain it on $C_i$, even if more attractive
elements could exist on this point of view. 
The practical identification of these absolute elements
has been performed in many real applications, see for example
\cite{baencova99,bava97,bava99}. 

In multicriteria decision aid, after the structuring phase comes
the evaluation phase, in which for each point of view $C_i$,
\emph{intra-criterion} information is gathered (i.e. attractiveness
for the DM of the elements of $X_i$ with respect to point of view
$C_i$), and also, according to an aggregation model chosen in agreement
with the DM, \emph{inter-criteria} information. This information,
which aims at determining the parameters of the chosen aggregation
model, generally consists in some information on the attractiveness for
the DM of some particular elements of $X=X_1\times\cdots\times
X_n$. These elements are selected so as to enable the resolution of
some equation system, whose variables are precisely the unknown
parameters of the aggregation model. 

In this paper, of which aim is primarily theoretical, we adopt with
respect to the classical approach described above, a rather converse
attitude. Specifically, we do not suppose to have beforehand a given
aggregation model, but rather to have some information concerning the
attractiveness for the DM of a particular collection of elements of
$X$. Then we study how to extend this information on the preference of
the DM to all elements of $X$. This kind of problem can be called an
\emph{identification of an aggregation model} which is compatible with
available information. 

The paper is organized as follows. In section \ref{sec:baas}, we
introduce the basic assumptions we make concerning the knowledge on
the attractiveness  for the DM of particular elements of $X$. Section
\ref{sec:incr} shows that this kind of information is compatible with
the existence of some interaction phenomena between points of view,
and introduces some definitions related to the concept of
interaction. The problem of extending the information on preferences
assumed to be known on a subpart of $X$, to the whole set $X$, is
addressed in section \ref{sec:defr}, and appears to be the problem of
identifying an aggregation model  compatible  with given
intra-criterion and inter-criteria information. In section
\ref{sec:exps}, we show that this problem amounts to define the
extension of a given pseudo-Boolean function, and we introduce some
possible extensions, which we relate to already known models in the
literature (section \ref{sec:liex}). Section \ref{sec:prex} briefly studies the
properties of these models, and concludes about their usefulness in this
context. In section 8, we show an equivalent set of axioms for our
construction, and in section 9, we address the question of unicity of the
solution. 

This paper does not deal with the practical aspects of the methodology
we are proposing, i.e. how to obtain the necessary information for
building the aggregation model. However, the MACBETH approach
\cite{bava94} could be most useful for extracting the information from
the DM.

Lastly, we want to mention that one of the reasons which have
motivated this research is the recent development of multicriteria
methods based on capacities and the Choquet integral \cite{cho53},
which seems to open new horizons \cite{gra95a,grbala96,grro98}. In a
sense, this paper aims at giving a theoretical foundation of this type
of approach in the framework of multicriteria decision making.

\section{Basic assumptions}
\label{sec:baas}
We present two basic assumptions, which are the starting point of our
construction. We denote the index set of criteria by
$N=\{1,\ldots,n\}$. Considering two acts $x,y\in X$, and $A\subset N$, we will
often use the notation $(x_A,y_{A^c})$ to denote the compound act $z$ where
$z_i = x_i$ if $i\in A$ and $y_i$ otherwise. $\wedge,\vee$ denote respectively
min and max operators. 

\subsection{Intra-criterion assumption}
\label{sec:intra}
We consider the particular subsets $X\rfloor_i$, $i=1,\ldots,n$, of
$X$, which are defined by:
\[
X\rfloor_i=
\{(\bzero_1,\ldots,\bzero_{i-1},x_i,\bzero_{i+1},\ldots,\bzero_n)|
x_i\in X_i\}.
\]
Using our convention, acts in $X\rfloor_i$ are denoted more simply by $(x_i, \bzero_{\{i\}^c})$.

We assume to have an interval scale denoted $v_i$ on each
$X\rfloor_i$, which quantifies the attractiveness for the DM of the
elements of $X\rfloor_i$ (assumption A1). In order to simplify the notation, we
denote for all $i\in N$, $u_i:X_i\longrightarrow \mathbb{R}$, $x_i\mapsto u_i(x_i)
= v_i(x_i, \bzero_{\{i\}^c})$. Thus, assumption A1 means exactly the following:
\begin{itemize} 
\item [(A1.1)] $\forall x_i,y_i\in X_i$, $u_i(x_i) \geq u_i(y_i)$ if
and only if for the decision maker
$(x_i,\bzero_{\{i\}^c})$ is
at least as attractive as
$(y_i,\bzero_{\{i\}^c})$.
\item [(A1.2)] $\forall x_i,y_i,z_i,w_i\in X_i$, such that
$u_i(x_i)>u_i(y_i)$ and $u_i(w_i)>u_i(z_i)$, we have
\[
\frac{u_i(x_i) - u_i(y_i)}{u_i(w_i) - u_i(z_i)} = k, \ \ \ k\in \mathbb{R}^+
\]
if and only if the difference of attractiveness that the DM feels between
$(x_i,\bzero_{\{i\}^c})$ and $(y_i,\bzero_{\{i\}^c})$ is equal to $k$ times the
difference of attractiveness between $(w_i,\bzero_{\{i\}^c})$ and
$(z_i,\bzero_{\{i\}^c})$.
\end{itemize}

We recognize here information concerning the intra-criterion preferences (i.e.
the attractiveness of elements of $X_i$ relatively to $C_i$), hence the name of
the assumption, which is a classical type of information in multicriteria
decision aid. Observe however that our presentation avoids the introduction of
any independence assumption (preferential or cardinal). This is possible since
we have introduced in every set $X_i$ an element $\bzero_i$ with an absolute
meaning in terms of attractiveness. This strong meaning allows us to fix
naturally $u_i(\bzero_i) = 0$,\footnote{which is technically always possible,
since an interval scale is defined up to a positive affine transformation
$\phi(z) = \alpha z + \beta$, $\alpha>0$, which means that we have two degrees
of freedom.} $i=1,\ldots,n$, and thus to consider $u_i$ as a ratio scale on
$X_i$. We can also take advantage of the remaining degree of freedom to fix the
value of $u_i(\bone_i)$. Contrarily to the case of $u_i(\bzero_i)$, no
particular value, provided it is positive, is mandatory here. However, since
all elements $\bone_i$, $i=1,\ldots,n$ have all the same absolute meaning, we
have to choose for $u_i(\bone_i)$ the same numerical value for all
$i\in\{1,\ldots,n\}$, which implies that the only admissible transformations of
the scales $u_i$, $i\in N$, are of the form $\phi(u_i) = \alpha\cdot u_i$,
where $\alpha>0$ does not depend on $i$. Thanks to the elements $\bzero_i$ and
$\bone_i$, the interval scales $u_i$ become thus \emph{commensurable ratio
scales}. In the sequel, we take as a convention $u_i(\bone_i) = 1$, for
$i=1,\ldots,n$.

\subsection{Inter-criteria assumption}
\label{sec:inter}
We consider now another subset of $X$, denoted $X\rceil_{\{0,1\}}$,
containing the following elements:
\[
X\rceil_{\{0,1\}} := \{(\bone_A,\bzero_{A^c})|A\subset N\},
\]
where $(\bone_A,\bzero_{A^c})$ denotes an act $(x_1,\ldots,x_n)$
with $x_i=\bone_i$ if $i\in A$ and $x_i=\bzero_i$ otherwise, following our convention.

We assume to have an interval scale $u_{\{0,1\}}$ on $X\rceil_{\{0,1\}}$,
quantifying the attractiveness for the DM of all elements in this set
(assumption A2). This means that:
\begin{itemize}
\item [(A2.1)] for all $A,B\subset N$, $u_{\{0,1\}}(\bone_A,\bzero_{A^c}) \geq
u_{\{0,1\}}(\bone_B,\bzero_{B^c})$ if and only if for the DM
$(\bone_A,\bzero_{A^c})$ is at least as attractive as
$(\bone_B,\bzero_{B^c})$.
\item [(A2.2)] for all $A,B,C,D\subset N$ such that
$u_{\{0,1\}}(\bone_A,\bzero_{A^c}) > u_{\{0,1\}}(\bone_B,\bzero_{B^c})$ and
$u_{\{0,1\}}(\bone_C,\bzero_{C^c}) > u_{\{0,1\}}(\bone_D,\bzero_{D^c})$, we have
\[
\frac{u_{\{0,1\}}(\bone_A,\bzero_{A^c}) - u_{\{0,1\}}(\bone_B,\bzero_{B^c})}{u_{\{0,1\}}(\bone_C,
\bzero_{C^c}) - u_{\{0,1\}}(\bone_D,\bzero_{D^c})} = k, \ \ \ k\in \mathbb{R}^+
\]
if and only if the difference of attractiveness felt by the DM between
$(\bone_A,\bzero_{A^c})$ and $(\bone_B,\bzero_{B^c})$ is $k$ times
the difference of attractiveness between $(\bone_C,\bzero_{C^c})$ and
$(\bone_D,\bzero_{D^c})$.
\end{itemize}
As we did for the case of intra-criterion information, we use the two
available degrees of freedom of an interval scale to fix:
\begin{align*}
u_{\{0,1\}}(\bone_{\emptyset},\bzero_{N})  = u_{\{0,1\}}(\bzero_1,\ldots,\bzero_n):= & 0\\
u_{\{0,1\}}(\bone_{N},\bzero_{\emptyset})  = u_{\{0,1\}}(\bone_1,\ldots,\bone_n):= & 1.
\end{align*}
Having in mind the meaning of $\bzero_i$, $i=1,\ldots,n$, it is natural to
impose $ u_{\{0,1\}}(\bzero_1,\ldots,\bzero_n)= 0$. The scale $u_{\{0,1\}}$ is
then a ratio scale. Let us point out that any strictly positive value could
have been used  instead of 1 for the value of
$u_{\{0,1\}}(\bone_1,\ldots,\bone_n)$. However, it is convenient to impose that
the value of $u_{\{0,1\}}(\bone_1,\ldots,\bone_n)$ is equal to the common value
chosen for the $u_i(\bone_i)$. 

At this point, let us remark that both $u_i(\bone_i)$ and
$u_{\{0,1\}}(\bone_i,\bzero_{\{i\}^c})$ quantify the attractiveness of act
$(\bone_i,\bzero_{\{i\}^c})$ for the DM, however their values are on different
ratio scales, but  with the same 0 since
$u_i(\bzero_i)=u_{\{0,1\}}(\bzero_1,\ldots,\bzero_n)=0$. This means that there
exists $K_i>0$ such that $u_{\{0,1\}}(\bone_i,\bzero_{\{i\}^c}) = K_i u_i(\bone_i)$. An important consequence of this fact is that, in order to
have compatibility between these scales (and hence between assumptions A1 and A2), we must have 
\[ u_{\{0,1\}}(\bone_{i},\bzero_{\{i\}^c}) >
u_{\{0,1\}}(\bzero_1,\ldots,\bzero_n) = 0, \ \ \ \forall i, 
\] 
otherwise no constant $K_i$ could exist.  This is not
restrictive on a practical point of view as soon as each point of view really
corresponds to a concern of the DM.

We suppose in addition that whenever $A\subset B$, the act $(\bone_B,\bzero_{B^c})$ is at least as attractive as $(\bone_A,\bzero_{A^c})$, which is also
a natural requirement.

Under these conditions, and introducing the set function $\mu:\mathcal{P}(N)\longrightarrow [0,1]$ by
\begin{equation}
  \label{eq:mu}
\mu(A) := u_{\{0,1\}}(\bone_A,\bzero_{A^c})
\end{equation}
we have defined a non-additive measure, or \textrm{\emph{fuzzy measure}},
\cite{sug74} or \textrm{\emph{capacity}} \cite{cho53}, with the additional
requirement that $\mu(\{i\}) > 0$. Indeed, a capacity is any non negative set
function such that $\mu(\emptyset) = 0$, $\mu(N)=1$, and $\mu(A)\leq\mu(B)$
whenever $A\subset B$.

\section{Interaction among criteria}
\label{sec:incr}
\begin{figure}[htb]
  \begin{center}
    $
    \epsfxsize=13.5cm 
    \epsfbox{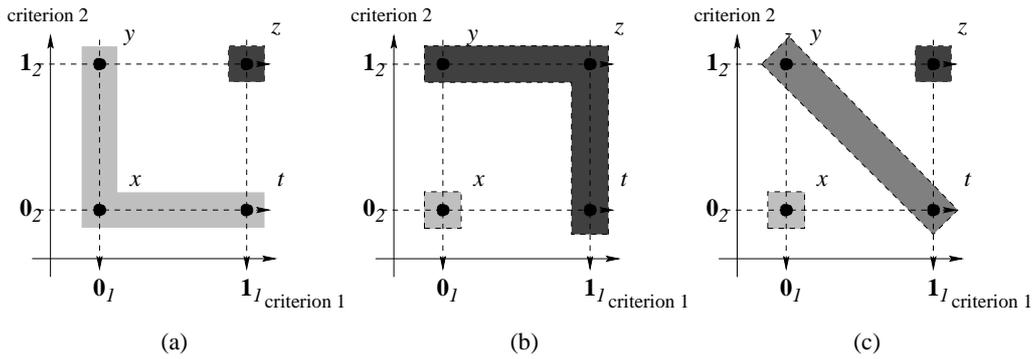} 
    $
  \end{center}
\caption{Different cases of interaction}
\label{fig:incr}
\end{figure}

Except the natural assumptions above for $\mu$ (monotonicity and $\mu(i)>0$ for
all $i\in N$), no restriction exists on $\mu$. Let us take 2 criteria to show
the range of decision behaviours we can obtain with capacities. We suppose in
addition that $\mu(\{1\}) = \mu(\{2\})$, which means that the DM is indifferent
between $(\bone_1,\bzero_2)$ and $(\bzero_1,\bone_2)$ (i.e. equal importance of
criteria, see section \ref{sec:defr}), and consider 4 acts $x,y,z,t$ such that
(see figure \ref{fig:incr}):
\begin{itemize}
\item $x = (\bzero_1, \bzero_2)$
\item $y = (\bzero_1, \bone_2)$
\item $z = (\bone_1,  \bone_2)$
\item $t = (\bone_1,  \bzero_2)$
\end{itemize}
Clearly, $z$ is more attractive than $x$ (written $z\succ x$), but preferences
over other pairs may depend on the decision maker. Due to the definition of
capacities, we can range from the two extremal following situations (recall
that $\mu(\{1,2\}) = 1$ is fixed):
\begin{description}
\item[extremal situation 1 (lower bound):] we put $\mu(\{1\})=\mu(\{2\}) = 0$,
  which is equivalent to the preferences $x\sim y\sim t$, where $\sim$ means
  indifference (figure \ref{fig:incr}, left).
\item[extremal situation 2 (upper bound):] we put $\mu(\{1\})= \mu(\{2\}) = 1$,
  which is equivalent to the preferences $y\sim z\sim t$ (figure \ref{fig:incr},
  middle).
\end{description}
Note that the first bound cannot be reached due to the condition
$\mu(i)>0$. The exact intermediate situation is $\mu(\{1\})=\mu(\{2\}) = 1/2$,
meaning that $z\succ y\sim t\succ x$ (figure \ref{fig:incr}, right), and the
difference of attractiveness between $x$ and $y$, $t$ respectively is the same
than between $z$ and $y$, $t$ respectively.

The first case corresponds to a situation where the criteria are
\emph{complementary}, since both have  to be satisfactory in order to get a
satisfactory act. Otherwise said, the DM makes a conjunctive aggregation. We
say that in such a case, which can be characterized by the fact that
$\mu(\{1,2\})> \mu(\{1\}) + \mu(\{2\})$, there is a \emph{positive interaction}
between criteria.

The second case corresponds to a situation where the criteria are
\emph{substitutive}, since only one has to be satisfactory in order to get a
satisfactory act. Here, the DM aggregates disjunctively. We say that in such a
case, which can be characterized by the fact that $\mu(\{1,2\})< \mu(\{1\}) +
\mu(\{2\})$, there is a \emph{negative interaction} between criteria.

In the third case, where we have $\mu(\{1,2\})= \mu(\{1\}) + \mu(\{2\})$, we
say that there is \emph{no interaction} among criteria, they are
\emph{non interactive}.

The information we assume to have at hand concerning the attractiveness
of acts for the DM is thus perfectly compatible with the interaction situations
between criteria, situations which are worth to consider on a practical point
of view, but up to now very little studied.

In the above simple example, we had only 2 criteria. In the general case, we
use the following definition proposed by Murofushi and Soneda \cite{muso93}.
\begin{define}
The \emph{interaction index} between criteria $i$ and $j$ is given by:
\begin{align} 
I_{ij} := & \sum_{K\subset N\setminus
\{i,j\}}\frac{(n-|K|-2)!|K|!}{(n-1)!}[\mu(K\cup\{i,j\})-\mu(K\cup\{i\})-
\nonumber \\
        & \mu(K\cup\{j\})+\mu(K)].
\end{align}
\end{define}
The definition of this index has been extended to any coalition $A\subset N$ of
criteria by Grabisch \cite{gra96f}:
\begin{equation} 
\label{eq:int}
I(A):=\sum_{B\subset N\setminus A}\frac{(n-|B|-|A|)!|B|!}{(n-|A|+1)!}\sum_{K\subset
A}(-1)^{|A|-|K|}\mu(K\cup B),\forall A\subset N.
\end{equation}
We have $ I_{ij} = I(\{i,j\})$.
When $A=\{i\}$, $I(\{i\})$ is nothing else than the Shapley value of game
theory \cite{sha53}. 
Properties of this set function has been studied and related to the
M\"obius transform \cite{degr96}. Also, $I$ has been characterized
axiomatically by Grabisch and Roubens \cite{grro97a}, in a way similar
to the Shapley index. Note that $I_{ij}>0$ (resp. $<0, =0$) for
complementary (resp. substitutive, non interactive) criteria.

\section{Constructing the model}
\label{sec:defr}
We will only consider in this paper the general type of aggregation model
introduced by Krantz \emph{et al.} \cite[Chap. 7]{krlusutv71}:
\begin{quote}
Act $x=(x_1,\ldots,x_n)$ is at least as attractive as act
$y=(y_1,\ldots,y_n)$ if and only if 
\[
F(u_1(x_1),\ldots,u_n(x_n))\geq F(u_1(y_1),\ldots,u_n(y_n)),
\]
\end{quote}
where the aggregation function $F:\mathbb{R}^n\longrightarrow \mathbb{R}$ is
strictly increasing in all its arguments. 

Indeed, this type of model is largely used, and has the advantage of being
rather general, and to lead to a complete and transitive preference relation on
$X$.  

The central question  we deal with in this paper is the identification of an
aggregation function $F$ which is compatible with intra-criterion and
inter-criteria information defined by assumptions A1 and A2, and satisfies
natural conditions. Specifically, we are looking for a mapping
$F:\mathbb{R}^n\longrightarrow \mathbb{R}$ of the form
\[
F(u_1(x_1),\ldots,u_n(x_n)) = u(x_1,\ldots,x_n) 
\]
satisfying the following requirements (in
which the presence of $\alpha$ is due to the fact that the $u_i$ are
commensurable ratio scales):
\begin{description}
\item [(i) compatibility with intra-criteria information] (assumption A1)
  \begin{itemize}
  \item $\forall i\in N$ and $\forall x_i,y_i\in X_i$,
\[
u_i(x_i)\geq u_i(y_i) \Leftrightarrow  u(x_i,\bzero_{\{i\}^c}) \geq
u(y_i,\bzero_{\{i\}^c})
\]
which becomes, in terms of $F$ (due to the consequences of assumption A1 on the scale):
\begin{multline}
\label{eq:a11}
u_i(x_i)\geq u_i(y_i) \Leftrightarrow \\
F(0,\ldots,0,\alpha u_i(x_i),0,\ldots,0)\geq F(0,\ldots,0,\alpha u_i(y_i),0,\ldots,0)
\end{multline}
for all $\alpha>0$. In fact, the constant $\alpha$ here is useless, since for
  any $\alpha>0$, $u_i(x_i)\geq u_i(y_i)\Leftrightarrow \alpha u_i(x_i)\geq
  \alpha u_i(y_i)$.
  \item $\forall i\in N$ and $\forall w_i,x_i,y_i,z_i$ such that $u_i(w_i)>u_i(x_i)$ and $u_i(y_i)>u_i(z_i)$,
\[
\frac{u(w_i,\bzero_{\{i\}^c}) -
 u(x_i,\bzero_{\{i\}^c}) }
{u(y_i,\bzero_{\{i\}^c}) -
u(z_i,\bzero_{\{i\}^c})  } =
\frac{u_i(w_i) - u_i(x_i)}{u_i(y_i) - u_i(z_i)}
\]
which becomes in terms of $F$:
\begin{multline}
  \label{eq:4i}
\frac{F(0,\ldots,0,\alpha u_i(w_i),0,\ldots,0) - F(0,\ldots,0,\alpha u_i(x_i),0,\ldots,0)}{F(0,\ldots,0,\alpha u_i(y_i),0,\ldots,0) - F(0,\ldots,0,\alpha u_i(z_i),0,\ldots,0)} =\\
 \frac{u_i(w_i) - u_i(x_i)}{u_i(y_i) - u_i(z_i)}
\end{multline}
for all $\alpha>0$.  
  \end{itemize}
\item [(ii) compatibility with inter-criteria information] (assumption A2)
  \begin{itemize}
  \item $\forall A,B\subset N$, we have
\[
u_{\{0,1\}}(\bone_A,\bzero_{A^c}) \geq u_{\{0,1\}}(\bone_B,\bzero_{B^c})
\Leftrightarrow u(\bone_A,\bzero_{A^c}) \geq u(\bone_B,\bzero_{B^c})
\]
which becomes, in terms of $F$:
\[
u_{\{0,1\}}(\bone_A,\bzero_{A^c}) \geq u_{\{0,1\}}(\bone_B,\bzero_{B^c})
\Leftrightarrow
 F(\alpha 1_A, 0_{A^c}) \geq F(\alpha 1_B, 0_{B^c})
\]
for all $\alpha>0$, where for any $A\subset N$, $(1_A,0_{A^c})$ is the vector
whose component   $x_i$ is 1 whenever $i\in A$, and 0 otherwise.
  \item $\forall A,B,C,D\subset N$, with $u_{\{0,1\}}(\bone_A,\bzero_{A^c}) >u_{\{0,1\}}(\bone_B,\bzero_{B^c})$ and\linebreak $u_{\{0,1\}}(\bone_C,\bzero_{C^c}) >u_{\{0,1\}}(\bone_D,\bzero_{D^c})$, we have:
\[
\frac{u(\bone_A,\bzero_{A^c}) - u(\bone_B,\bzero_{B^c})}
{u(\bone_C,\bzero_{C^c}) - u(\bone_D,\bzero_{D^c})} =
\frac{u_{\{0,1\}}(\bone_A,\bzero_{A^c}) - u_{\{0,1\}}(\bone_B,\bzero_{B^c})}
{u_{\{0,1\}}(\bone_C,\bzero_{C^c}) - u_{\{0,1\}}(\bone_D,\bzero_{D^c})}
\]
which becomes, in terms of $F$:
\begin{equation}
  \label{eq:1}
\frac{F(\alpha 1_A, 0_{A^c}) - F(\alpha 1_B, 0_{B^c})}{F(\alpha 1_C, 0_{C^c}) -
F(\alpha 1_D, 0_{D^c})} =
 \frac{u_{\{0,1\}}(\bone_A,\bzero_{A^c}) - u_{\{0,1\}}(\bone_B,\bzero_{B^c})}
{u_{\{0,1\}}(\bone_C,\bzero_{C^c}) - u_{\{0,1\}}(\bone_D,\bzero_{D^c})}
\end{equation}
for all $\alpha>0$. 
  \end{itemize}
\item [(iii) conditions related to absolute information]\mbox{}\\
We impose that scales $u$ and $u_{\{0,1\}}$ coincide on particular acts
corresponding to absolute information, namely:
   \begin{itemize}
   \item $u(\bzero_1,\ldots,\bzero_n)=
u_{\{0,1\}}(\bzero_1,\ldots,\bzero_n):=0$, which leads to\linebreak $F(0,\ldots,0) = 0$.
   \item $u(\bone_1,\ldots,\bone_n)=
u_{\{0,1\}}(\bone_1,\ldots,\bone_n):=1$, which leads to\linebreak
$F(1,\ldots,1) = 1$. However, remember that the choice of value ``1'' was
arbitrary when building scales $u_i$ and $u_{\{0,1\}}$, and any positive
constant $\alpha$ can do. Hence, we should satisfy more generally \linebreak $F(\alpha,
\ldots,\alpha) = \alpha$, $\forall \alpha>0$. 
   \end{itemize}
\item [(iv) monotonicity of $F$.] This property is a fundamental requirement for
any aggregation function:
\begin{multline*}
\forall (t_1,\ldots,t_n), \forall (t'_1,\ldots,t'_n)\in \mathbb{R}^n, \\
t'_i\geq t_i, i=1,\ldots,n \Rightarrow F(t'_1,\ldots,t'_n)\geq
F(t_1,\ldots,t_n).  
\end{multline*}
The monotonicity is strict if all inequalities are strict. Remark that
monotonicity entails the first condition of \textbf{(i)}, namely formula (\ref{eq:a11}). 
\end{description}
Let us remark that, as suggested in \textbf{(iv)} above, that $F$ can be
viewed as an aggregation function, and thus our problem amounts to the search of
an aggregation model which is compatible with intra- and inter-criteria
information defined by assumptions A1 and A2.

At this point, let us make two remarks.
\begin{itemize}
\item the reader may wonder about the very specific form of inter-criteria
  information asked for, that is, attractiveness of acts of the form
  $(\bone_A,\bzero_{A^c})$. These acts present the double advantage to be non
  related with real acts, which permits to avoid any emotional answer from the
  DM, and to have, taking into account the definition of $\bzero_i$ and
  $\bone_i$, a very clear meaning, and consequently, to be very well perceived
  and understood.

  They are currently used in real world applications of the MACBETH approach
  \cite{baencova99,bava97,bava99} . Until now, these applications were done in
  the framework of an additive aggregation model. In such a case, only acts of
  the form $(\bone_i,\bzero_{\{i\}^c})$ have to be introduced.
  
  What we are doing here is merely a generalization, considering not only
  single criteria, but any \emph{coalition} of criteria. This natural
  generalization from singletons to subsets is indeed the key to the modelling
  of interaction, as explained in section \ref{sec:incr}. In this sense, the
  global utility $u(\bone_A,\bzero_{A^c})$, which is a capacity (see section
  \ref{sec:inter}), could represent the importance of coalition $A$ to make
  decision.

  It must be noted, however, that we assume that all acts
  $(\bone_A,\bzero_{A^c})$ are at least conceivable, i.e. the conjunction of
  attributes in $A$ being ``good'' and the other ones being ``neutral'', do not
  lead to a logical impossibility or contradiction. This could happen when
  some attributes are strongly correlated, a situation which should be avoided
  in multicriteria decision making.

\item it can be observed that conditions \textbf{(ii)} and \textbf{(iii)}
  above entail that the function $F:\mathbb{R}^n\longrightarrow \mathbb{R}$ to be
  determined must coincide with $\mu$ on $\{0,1\}^n$, i.e.:
  \[
F(1_A,0_{A^c}) = \mu(A), \quad \forall A\subset N.
  \]
  Indeed, just consider equation (\ref{eq:1}) with $B=D=\emptyset, C=N$, and
  use  \textbf{(iii)}, and definition of $\mu$ (eq. (\ref{eq:mu})).
  
  Thus, $F$ must be an extension of $\mu$ on $\mathbb{R}^n$. In other words,
  \emph{the assignment of importance to coalitions is tightly linked with the
    evaluation function}. This fact is well known in the MCDM community (see
  e.g. Mousseau \cite{mou92}), but the argument above puts it more precisely.
  The next section addresses in full detail the problem of extending
  capacities. 
\end{itemize}

\section{Extension of pseudo-Boolean functions}
\label{sec:exps}
The problem of extending a capacity can be nicely formalized through the use of
pseudo-Boolean functions (see e.g. \cite{haho92}). 

Any function
$f:\{0,1\}^n\longrightarrow\mathbb{R}^{}$ is a said to be a
\emph{pseudo-Boolean function}. By making the usual bijection between
$\{0,1\}^n$ and $\mathcal{P}(N)$, it is clear that pseudo-Boolean
functions on $\{0,1\}^n$ coincide with real-valued set functions on $N$ (of
which capacities are a particular case). More
specifically, if we define for any subset $A\subset N$ the vector
$\delta_A=[\delta_A(1) \cdots \delta_A(n)]$ in $\{0,1\}^n$ by $\delta_A(i) = 1$ if
$i\in A$, and 0 otherwise, then for any set function $v$ we can define
its associated pseudo-Boolean function $f$ by
\[
f(\delta_A) := v(A), \ \ \ \forall A\subset N,
\]
and reciprocally. It has been shown by Hammer and Rudeanu
\cite{haru68} that any pseudo-Boolean function can be written in a
multilinear form:
\begin{equation} 
\label{eq:mlf}
f(t) = \sum_{A\subset N}m(A)\cdot\prod_{i\in A}t_i, \ \ \ \forall
t\in\{0,1\}^n.
\end{equation}
$m(A)$ corresponds to the M\"obius transform (see e.g. Rota
\cite{rot64}) of
$v$, associated to $f$, which is defined by:
\begin{equation} 
m(A) = \sum_{B\subset A} (-1)^{|A\setminus B|}v(B).
\end{equation}
Reciprocally, $v$ can be recovered from the M\"obius transform by
\begin{equation}
\label{eq:mobcar}
v(A) = \sum_{B\subset A}m(B).
\end{equation}
If necessary, we write $m^v$ for the M\"obius transform of $v$. 
Note that (\ref{eq:mlf}) can be put in an equivalent form, which is
\begin{equation}
\label{eq:mmf}
f(t) = \sum_{A\subset N} m(A)\cdot \bigwedge_{i\in A}t_i, \ \ \ \forall
t\in\{0,1\}^n.
\end{equation}
More generally, the product can be replaced by any operator $\boxdot$ on
$[0,1]^n$ coinciding with the product on $\{0,1\}^n$, such as t-norms
\cite{scsk83} (see e.g. \cite{foro94} for a survey on this topic, and
\cite{klmepa00} for a complete treatment). We recall that a t-norm is a binary
operator $T$ on $[0,1]$ which is commutative, associative, non decreasing in
each place, and such that $T(x,1) = x$, for all $x\in [0,1]$. Associativity
permits to unambiguously define t-norms for more than 2 arguments.

These are not the only ways to write pseudo-Boolean functions. When $v$ is a
capacity, it is possible 
to replace the sum by $\vee$, as the following formula shows \cite{gra97a}:
\begin{equation}
\label{eq:sug}
f(t) = \bigvee_{A\subset N} m_\vee (A)\wedge \left(\bigwedge_{i\in A} t_i\right).
\end{equation}
The quantity $m_\vee$ is called the \emph{ordinal M\"obius transform}, and is
related to $v$ by $m_\vee(A) = v(A)$ whenever $v(A)>v(A\setminus i)$ for all
$i\in A$, and 0 otherwise. Note that conversely we have (compare with (\ref{eq:mobcar})):
\begin{equation}
\label{eq:mobord}
v(A) =\bigvee_{B\subset A} m_\vee(B), \forall A\subset N.
\end{equation}

In the sequel, we focus on formulas (\ref{eq:mlf}) and (\ref{eq:mmf}). We will
come back on alternatives to these formulas  in section \ref{sec:8}.

In order to extend $f$ to $\mathbb{R}^n$, which is necessary in our framework
since the DM can judge that an element $(x_i,\bzero_{\{i\}^c})$ is less
attractive than $(\bzero_1,\ldots,\bzero_n)$ (in that case $u_i(x_i)<0$), two
immediate extensions come from (\ref{eq:mlf}) and (\ref{eq:mmf}), where we
simply use any $t\in\mathbb{R}^n$ instead of
$\{0,1\}^n$. We will denote them 
\begin{equation}
  \label{eq:mul}
f^\Pi(t) := \sum_{A\subset N}m(A)\cdot\prod_{i\in A}t_i, \ \ \ \forall
t\in\mathbb{R}^n, 
\end{equation}
\begin{equation}
\label{eq:lov}
f^\wedge(t) := \sum_{A\subset N} m(A)\cdot\bigwedge_{i\in A} t_i, \ \ \
\forall t\in\mathbb{R}^n.
\end{equation}
However a second way can be obtained by considering the fact that any real
number $t$ can be written under the form $t=t^+-t^-$, where $t^+=t\vee 0$, and
$t^-=-t\vee 0$. If, by analogy with this remark, we replace $\prod_i t_i$ by
$\prod_i t^+_i - \prod_i t^-_i$, and similarly with $\bigwedge$, we obtain two
new extensions:
\begin{equation}
  \label{eq:muls}
f^{\Pi\pm}(t) := \sum_{A\subset N}m(A)\left[\prod_{i\in
A}t_i^+ - \prod_{i\in A}t_i^-\right], \ \ \ \forall
t\in\mathbb{R}^n, 
\end{equation}
\begin{equation}
\label{eq:lovs}
f^{\wedge\pm}(t) := \sum_{A\subset N} m(A)\left[ \bigwedge_{i\in A}
t_i^+ - \bigwedge_{i\in A} t_i^- \right], \ \ \
\forall t\in\mathbb{R}^n.
\end{equation}
These are not the only possible extensions. In fact, nothing prevents us to
introduce for the negative part another capacity, e.g. equation (\ref{eq:lovs})
could become:
\begin{equation}
\label{eq:lovss}
f^{\wedge\pm}_{12}(t) := \sum_{A\subset N} m_1(A)\cdot \bigwedge_{i\in A}
t_i^+ -  \sum_{A\subset N} m_2(A)\cdot\bigwedge_{i\in A} t_i^-, \ \ \
\forall t\in\mathbb{R}^n.
\end{equation}
However, we will not consider this possibility in the subsequent
development, except in section 9 where the question of unicity is addressed. 
In the next sections we investigate whether  extensions (\ref{eq:mul}) to
(\ref{eq:lovs}) are related to
known models of aggregation, and which one satisfy the requirements (i) to (iv)
 introduced in section \ref{sec:defr}, and can be thus used as an aggregation
 function in our case.

\section{Link with existing models}
\label{sec:liex}
We introduce the Choquet integral with respect to a capacity, which has
been introduced as an aggregation operator by Grabisch \cite{gra94b,gra95a}.
Let $\mu$ be a capacity on $N$, and $t=(t_1,\ldots,t_n)\in
(\mathbb{R}^+)^n$. 
The \emph{Choquet integral} of $t$ with respect to $\mu$ is defined by \cite{musu89}:
\begin{equation}
\label{eq:cho1}
\mathcal{C}_\mu(t) = \sum_{i=1}^n (t_{(i)} - t_{(i-1)})\mu(\{(i),\ldots,(n)\})
\end{equation}
where $\cdot_{(i)}$ indicates a permutation on $N$ so that $t_{(1)}\leq t_{(2)}
\leq \cdots\leq t_{(n)}$, and $t_{(0)} :=0$ by convention. It can be shown
that the Choquet integral can be written as follows:
\begin{equation}
  \label{eq:mob}
\mathcal{C}_\mu(t) = \sum_{A\subset N} m(A) \bigwedge_{i\in A}t_i, \quad\forall
t\in (\mathbb{R}^+)^n
\end{equation}
where $m$ denotes the M\"obius transform of $\mu$.
This result has been shown
by Chateauneuf and Jaffray \cite{chja89} (also by Walley \cite{wal81}),
extending Dempster's result \cite{dem67}.

We are now ready to relate previous extensions to known aggregation models.
\begin{itemize}
\item the extension $f^\Pi$ is known in multiattribute utility theory as the
  \emph{multilinear model} \cite{kera76}, which we denote by MLE. Note that our
  presentation gives a meaning to the coefficients of the polynom, since they
  are the M\"obius transform of the underlying capacity defined by $\mu(A) =
  u(\bone_A, \bzero_{A^c})$, for all $A\subset N$. Up to now, no clear
  interpretation of these coefficients were given.
\item concerning $f^{\Pi\pm}$, to
our knowledge, it does not correspond to anything known in the literature. We
will denote it by SMLE (symmetric MLE).
\item considering $f^\wedge$ restricted to $(\mathbb{R}^+)^n$, it appears due to
the above result (\ref{eq:mob}) that $f^\wedge$ is the Choquet integral of $t$
with respect to $\mu$, where $\mu$ corresponds to $f$. This extension is also
known as the Lov\'asz extension of $f$ \cite{lov83,sin84}. At this point, let
us remark that the extension of the Choquet integral to negative arguments has
been considered by Denneberg \cite{den94}, who gives two possibilities:
\begin{enumerate} \item the symmetric extension
$\overset{\mathrm{S}}{\mathcal{C}}_\mu$ defined by 
\begin{equation}
\label{eq:11bis} \overset{\mathrm{S}}{\mathcal{C}}_\mu(t) =
\mathcal{C}_\mu(t^+) -\mathcal{C}_\mu(t^-), \quad \forall t\in \mathbb{R}^n.
\end{equation} 
\item the asymmetric extension
$\overset{\mathrm{AS}}{\mathcal{C}}_\mu$ defined by 
\begin{equation}
\label{eq:11ter}
\overset{\mathrm{AS}}{\mathcal{C}}_\mu(t) = \mathcal{C}_\mu(t^+)
-\mathcal{C}_{\bar{\mu}}(t^-), \quad \forall t\in \mathbb{R}^n, 
\end{equation}
where
$\bar{\mu}$ is the conjugate capacity defined by $\bar{\mu}(A):= \mu(N) -
\mu(A^c)$.  \end{enumerate} The first extension has been proposed first by
\Sipos{} \cite{sip79}, while the second one is considered as the classical
definition of the Choquet integral on real numbers. In the sequel, we will
denote the \Sipos{} integral by $\check{\mathcal{S}}_\mu$, while we keep
$\mathcal{C}_\mu$ for the (usual) Choquet integral.
\end{itemize}
The following proposition gives the expression of Choquet and \Sipos{}
integrals in terms of the M\"obius transform, and shows that $f^\wedge\equiv
\mathcal{C}_\mu$ and $f^{\wedge\pm}\equiv\check{\mathcal{S}}_\mu$.
\begin{prop}\label{prop:mob}
Let $\mu $ be a capacity. For any $t\in \mathbb{R}^{n}$, 
\begin{align}
\label{eq:1.18}
\mathcal{C}_\mu(t) & = \sum_{A\subset N} m(A)\bigwedge_{i\in A} t_i,\\
\label{eq:1.19}
\Sint_\mu(t) & = \sum_{A\subset N} m(A)\left[ \bigwedge_{i\in A}
t_i^+ - \bigwedge_{i\in A} t_i^- \right] \nonumber \\
& =\sum_{A\subset
N^+}m(A)\bigwedge_{i\in A}t_i + \sum_{A\subset N^-}m(A)\bigvee_{i\in A}t_i,
\end{align}
where $N^+:=\{i\in N|t_i\geq 0\}$ and $N^-=N\setminus N^+$.
\end{prop}

The proof is based on the following lemma, shown in \cite{gra98c}.
\begin{lemma}
Let $v$ be any set function such that $v(\emptyset)=0$, and consider its 
\emph{co-M\"obius} transform\footnote{Called ``commonality function'' by
Shafer \cite{sha76}.} \cite{gra97}, defined by:
\[
\check{m}^v(A) := \sum_{B\supset N\setminus A}(-1)^{n-|B|}v(B) =
\sum_{B\subset A}(-1)^{|B|} v(N\setminus B), \forall
A\subset N.
\]
Then, if $\bar{v}$ denotes the conjugate
set function:
\begin{equation} 
\label{eq:1.22}
\check{m}^{\bar{v}} (A)  =  (-1)^{|A|+1}m^v(A), \ \ \ \forall A\subset
N, A\neq\emptyset
\end{equation}
and for any $a\in (\mathbb{R}^+)^n$,
\begin{equation} 
\label{eq:1.23}
{\cal C}_v(a)  =  \sum_{A\subset N,A\neq\emptyset}
(-1)^{|A|+1}\check{m}^v(A) \bigvee_{i\in A}a_i.
\end{equation}
\end{lemma}
{\bf Proof of Prop. \ref{prop:mob}:} The case of \Sipos\ integral is clear from
(\ref{eq:lov}) and (\ref{eq:11bis}). For the case of
Choquet, the proof is based on the above lemma. Using (\ref{eq:lov}), we have:
\begin{align*}
\mathcal{C}_\mu(t^+) & = \sum_{A\subset N}m(A)\bigwedge_{i\in A}t_i^+\\
        & = \sum_{A\subset N, A\cap N^-=\emptyset}m(A)\bigwedge_{i\in A}t_i
\end{align*}
Also, using (\ref{eq:1.22}) and (\ref{eq:1.23}) and remarking that
$m(\emptyset)=0$, we get:
\begin{align*}
\mathcal{C}_{\bar{\mu}}(t^-) & = \sum_{A\subset N,
A\neq\emptyset}(-1)^{|A|+1}\check{m}^{\bar{\mu}}(A)\bigvee_{i\in
A}t_i^- \\ 
 & = \sum_{A\subset N}m(A)\bigvee_{i\in
A}t_i^-.
\end{align*}
Now
\[
\bigvee_{i\in A}t_i^- = \left\{ \begin{array}{ll}
                        -\bigwedge_{i\in A}t_i, & \text{ if } A\cap
                        N^-\neq \emptyset \\
                        0, & \text{otherwise}
                                \end{array} \right.
\]  
Thus
\[
\mathcal{C}_{\bar{\mu}}(t^-) = -\sum_{A\subset N, A\cap
N^-\neq\emptyset}m(A) \bigwedge_{i\in A}t_i
\]
so that
\[
\mathcal{C}_\mu(t) = \mathcal{C}_\mu(t^+) -
\mathcal{C}_{\bar{\mu}}(t^-) = \sum_{A\subset
N}m(A)\bigwedge_{i\in A}t_i.
\]
$\Box$

The next proposition gives the expression of Choquet and \Sipos{} integral
directly in terms of the capacity.
\begin{prop}
\label{pr:capa}
Let $\mu$ be a capacity. For any $t\in\mathbb{R}^n$,
\begin{align}
\mathcal{C}_{\mu}(t) & = t_{(1)} + \sum_{i=2}^n \pp{t_{(i)} - t_{(i-1)}}
\mu\pp{\pa{(i),\ldots,(n)}} \label{eq:choq}\\
\Sint_{\mu}(t)  = & \sum_{i=1}^{p-1} \pp{t_{(i)} - t_{(i+1)}}
         \mu\pp{\pa{(1),\ldots,(i)}}
        + t_{(p)}\mu\pp{\pa{(1),\ldots,(p)}}  \nonumber \\
    & + t_{(p+1)}\mu\pp{\pa{(p+1),\ldots,(n)}}
      + \sum_{i=p+2}^n \pp{t_{(i)} - t_{(i-1)}}
         \mu\pp{\pa{(i),\ldots,(n)}}
\label{ESi1}
\end{align}
where $\cdot_{(i)}$ indicates a permutation on $N$ so that
$t_{(1)}\leq t_{(2)} \leq \cdots \leq t_{(p)} < 0 \leq$ $t_{(p+1)}
   \leq \cdots \leq t_{(n)}$.
\end{prop}
\textbf{Proof:} from the definition (\ref{eq:cho1}), we have:
\[
\mathcal{C}_{\mu}(t)  = t_{(1)} + \sum_{i=2}^n \pp{t_{(i)} - t_{(i-1)}}
\mu\pp{\pa{(i),\ldots,(n)}}.
\]
Let
$t\in\mathbb{R}^{n}$. We  split $t$ into its positive and negative parts
$t^+, t^-$. Since
\[ \left\{ \begin{array}{l}
   (t^+)_{(1)}=(t^+)_{(2)}= \cdots =(t^+)_{(p)}=0 \\
   (t^+)_{(p+1)} = t_{(p+1)} \\
   \vdots \\
   (t^+)_{(n)} = t_{(n)}
   \end{array} \right.
\]
we have
\[ \mathcal{C}_{\mu}(t^+) = t_{(p+1)}\mu\pp{\pa{(p+1),\ldots,(n)}}
      + \sum_{i=p+2}^n \pp{t_{(i)} - t_{(i-1)}}
         \mu\pp{\pa{(i),\ldots,(n)}} \ .
\]
In the same way, one has
\[  \mathcal{C}_{\mu}(t^-) = -t_{(p)}\mu\pp{\pa{(p),\ldots,(1)}}
      - \sum_{i=1}^{p-1} \pp{t_{(i)} - t_{(i+1)}}
         \mu\pp{\pa{(i),\ldots,(1)}} \ .
\]
This gives the desired expression for \Sipos\ integral. The case of Choquet
integral proceeds similarly. $\Box$

Remarking that $\mathcal{C}_\mu(0)
=\Sint_\mu(0)$ for any capacity, we have from proposition~\ref{pr:capa}:
\begin{align}
\label{eq:1.20}
\mathcal{C}_\mu(-t) & = - \mathcal{C}_{\bar{\mu}}(t) \\
\label{eq:1.21}
\Sint_\mu(-t) & = - \Sint_\mu(t)
\end{align}
for any $t$ in $\mathbb{R}^{n}$, hence the terms asymmetric and symmetric.

In summary, three among the four extensions correspond to known models of
aggregation, even if contexts may differ.

\section{Properties of the extensions}
\label{sec:prex}
This section is devoted to the study of the four extensions, regarding the
properties requested in the construction of the aggregation model (section \ref{sec:defr}).

\paragraph{compatibility with intra-criterion information (assumption A1)}
Recalling that
$u_i(\boldsymbol{0}_i)=0$ $\forall i\in N$, and noting that $m(\{i\}) =
\mu(\{i\})$,  a straightforward
computation shows that for any $\alpha>0$:
\begin{align} 
\mathcal{C}_{\mu}(0,\ldots,0,\alpha u_i(x_i),0,\ldots,0) 
&   = \left\{ \begin{array}{ll}
   \alpha \mu(\{i\})u_i(x_i) & \mbox{ if }  x_i\succeq_i \boldsymbol{0}_i \\
   \alpha \bar{\mu}(\{i\})u_i(x_i) & \mbox{ if }  x_i\prec_i \boldsymbol{0}_i
   \end{array} \right. \\
\Sint_\mu(0,\ldots,0,\alpha u_i(x_i),0,\ldots,0)  & = \alpha \mu(\{i\})u_i(x_i)\\
\mathrm{MLE}_\mu(0,\ldots,0,\alpha u_i(x_i),0,\ldots,0) & = \alpha \mu(\{i\})u_i(x_i)\\
\mathrm{SMLE}_\mu(0,\ldots,0,\alpha u_i(x_i),0,\ldots,0) & = \alpha \mu(\{i\})u_i(x_i).
\end{align}
In the general case, we have $\mu(\pa{x_i}) \not=
\bar{\mu}(\pa{x_i})$. Thus there is an angular point around the
origin for the Choquet integral. The consequence is that equation
(\ref{eq:4i}), and hence assumption A1, are not satisfied by the Choquet integral in general. 

This curious property can be explained as follows. For the \Sipos\ integral,
the zero has a special role, since it is the zero of the ratio scale, and all
is symmetric with respect to this point. For the Choquet integral, the zero has
no special meaning, but observe that if $x_i\succeq \boldsymbol{0}_i\succeq
y_i$, the acts $(\boldsymbol{0}_1,\ldots
\boldsymbol{0}_{i-1},x_i,\boldsymbol{0}_{i+1},\ldots, \boldsymbol{0}_n)$ and
$(\boldsymbol{0}_1,\ldots \boldsymbol{0}_{i-1},y_i,\boldsymbol{0}_{i+1},\ldots,
\boldsymbol{0}_n)$ are not comonotonic, i.e. they induce a different ordering
of the integrand.

\paragraph{compatibility with inter-criteria information (assumption A2)}
It results from the definitions of $\mathcal{C}_\mu$, $\Sint_\mu$,
$\mathrm{MLE}_\mu$ and $\mathrm{SMLE}_\mu$ that, $\forall A\subset N$ and
$\forall \alpha>0$,
\begin{equation}
\label{eq:icmle}
\mathrm{MLE}_\mu(\alpha 1_A, 0_{A^c}) =
\mathrm{SMLE}_\mu(\alpha 1_A, 0_{A^c}) = \sum_{B\subset
A} m(B)\alpha^{|B|},
\end{equation}
and
\[
\mathcal{C}_{\mu}(\alpha 1_A, 0_{A^c}) =
\Sint_\mu(\alpha 1_A, 0_{A^c}) = \alpha\mu(A).
\]
Consequently, MLE and SMLE are inadequate for our model. 

\paragraph{use of absolute information}
Obviously any extension satisfies\linebreak $F(0,\ldots,0)=0$, and taking into account the
fact that $\mu(N)=1$, we have $\mathcal{C}_\mu(\alpha,\ldots,\alpha)  =
\Sint_\mu(\alpha, \ldots,\alpha)= \alpha$, for all $\alpha>0$. But from
(\ref{eq:icmle}), this property is not satisfied by MLE and SMLE.

\paragraph{Monotonicity}
It can be shown that, for any $t,t'\in \mathbb{R}^n$,
\begin{align}
t_i\leq t'_i, i=1,\ldots,n & \Rightarrow \mathcal{C}_\mu(t_1,\ldots,t_n)
\leq \mathcal{C}_\mu(t'_1,\ldots,t'_n)\\
t_i\leq t'_i, i=1,\ldots,n & \Rightarrow \Sint_\mu(t_1,\ldots,t_n)
\leq \Sint_\mu(t'_1,\ldots,t'_n).
\end{align}
This well-known result (see e.g. Denneberg \cite{den94}) comes from the fact
that for any  $t\in (\mathbb{R}^+)^n$, an equivalent form of (\ref{eq:cho1})
is:
\[
\mathcal{C}_\mu(t) = \sum_{i=1}^n t_{(i)}[\mu(\{(i),\ldots,(n)\}) - \mu(\{(i+1),\ldots,(n)\})].
\]
Monotonicity is immediate from the fact that $A\subset B$ implies
$\mu(A)\leq\mu(B)$. Now, for any $t\in \mathbb{R}^n$, monotonicity of the
Choquet and \Sipos\ integrals follow from equations (\ref{eq:11bis}) and
(\ref{eq:11ter}).  To obtain strict monotonicity, we need strict monotonicity
of the capacity, i.e. $A\subsetneqq B$ implies $\mu(A)<\mu(B)$.

It is easy to see from definition that MLE and SMLE are monotonic when
the coefficients $m(A)$ are all positive. But in general, the M\"obius
transform of a capacity is not always positive. To our knowledge,
there is no result in the general case. The following can be proven.
\begin{prop}
\label{prop:3}
For any $t\in [0,1]^n$, for any capacity $\mu$, $\mathrm{MLE}_\mu$ is
non decreasing with respect to $t_i$, $i=1,\ldots,n$. Strict
increasingness is ensured iff $\mu$ is strictly monotonic.
\end{prop}
\textbf{Proof:}  
We can express easily MLE with respect to $\mu$ (see Owen \cite{owe88}):
\[
\mathrm{MLE}_\mu(t) = \sum_{A\subset N} \left[\prod_{i\in
A}t_i\right]\left[\prod_{i\not\in A}(1-t_i)\right]\mu(A).
\]
Then we have, for any $t\in [0,1]^n$ and any $k\in N$:
\begin{align*}
\frac{\partial \mathrm{MLE}(t)}{\partial t_k}  = & \sum_{A\subset
N\setminus k} \left[\prod_{i\in
A}t_i\right]\left[\prod_{i\not\in A,i\neq k}(1-t_i)\right]\mu(A\cup
k)\\
& - \sum_{A\subset N\setminus k} \left[\prod_{i\in
A}t_i\right]\left[\prod_{i\not\in A,i\neq k}(1-t_i)\right]\mu(A)\\
= & \sum_{A\subset N\setminus k} \left[\prod_{i\in
A}t_i\right]\left[\prod_{i\not\in A,i\neq k}(1-t_i)\right](\mu(A\cup
k)- \mu(A)).
\end{align*}
Clearly, the expression is non negative (resp. positive) for any $k\in
N$ iff $\mu$ is monotonic (resp. strictly monotonic). $\Box$

The proof shows clearly that MLE could be non increasing when $t$ is
no more in $[0,1]^n$. Taking for example $n=2$, with
$\mu(\{1\})=\mu(\{2\})=0.9$, we have:
\begin{align*}
\mathrm{MLE}_\mu(1,1) & = 0.9 + 0.9 - 0.8 = 1\\
\mathrm{MLE}_\mu(3,3) & = (3)(0.9) + (3)(0.9) - (9)(0.8) = -1.8<
\mathrm{MLE}_\mu(1,1). 
\end{align*}
As a consequence, the use of MLE should be restricted to
criteria of which scores are limited to $[0,1]$, that is, \emph{unipolar
  bounded} criteria. Also, SMLE which differs from MLE only for negative
values, is clearly useless.

\medskip

\paragraph{Scale preservation}
Although this property is not required by our construction (but it somehow
underlies it in assumptions A1 and A2), it is
interesting to investigate whether the extensions satisfy it. 

The following is easy to prove.
\begin{itemize} 
\item [(C.1)] invariance to the same positive affine transformation
\[
\mathcal{C}_\mu(\alpha t_1+\beta,\ldots,\alpha t_n+\beta) =
\alpha\mathcal{C}_\mu(t_1,\ldots,t_n) + \beta, \ \ \ \forall
\alpha\geq 0, \forall \beta\in \mathbb{R}^{}.
\]
\item [(S.1)] homogeneity
\[
\Sint_\mu(\alpha t_1, \ldots, \alpha t_n) =
\alpha\Sint_\mu(t_1,\ldots,t_n), \forall \alpha\in \mathbb{R}^{}.
\]
\end{itemize}
As remarked by Sugeno and Murofushi \cite{sumu93}, this means that if
the scores $t_i$ are on commensurable interval scales, then the global score
computed by the Choquet integral is also on an interval scale
(i.e. relative position of the zero), and if the scores are on a ratio
scale, then the global score computed by the \Sipos{} integral is on a
ratio scale (absolute position of the zero).

By contrast, MLE and SMLE
neither preserve the interval nor the ratio scale, since they are not
homogeneous. Indeed, taking $n=2$ and any $\alpha\in\mathbb{R}^*$:
\begin{align*}
\mathrm{MLE}_\mu(\alpha t_1,\alpha t_2) & = m(\{1\})\alpha t_1 +
m(\{2\})\alpha t_2 +  m(\{1,2\})\alpha^2 t_1t_2 \\
        \neq \alpha \mathrm{MLE}_\mu(t_1,t_2).
\end{align*}
This is the reason why MLE and SMLE failed to fulfill assumption A2.  Note
however that MLE satisfies (\ref{eq:4i}) but not (\ref{eq:1}).

\medskip

As a conclusion, only the \Sipos\ integal among our four candidates can fit all
requirements of our construction.

\section{An equivalent  axiomatic}
\label{sec:8}
Our construction is based on a certain number of requirements for aggregation
function $F$, which we sum up below:
\begin{itemize}
\item restricted monotonicity (M1), coming from assumption A1:
\[
\forall i=1,\ldots,n, \forall a_i,a'_i\in \mathbb{R}, a_i\geq a'_i\Rightarrow F(a_i,0_{\{i\}^c}) \geq F(a'_i,0_{\{i\}^c})
\]
\item interval scale for intra-criterion information (A1):
\[
\frac{F(\alpha a_i,0_{\{i\}^c}) - F(\alpha b_i,0_{\{i\}^c}) }{F(\alpha
c_i,0_{\{i\}^c})-F(\alpha d_i,0_{\{i\}^c})} = \frac{a_i - b_i}{c_i - d_i}, \forall
\alpha>0, \forall a_i,b_i,c_i,d_i\in \mathbb{R}, c_i\neq d_i
\]
\item interval scale for inter-criteria information (A2):
\[
\frac{F(\alpha 1_A,0_{A^c}) - F(\alpha 1_B, 0_{B^c})}{F(\alpha 1_C,0_{C^c}) -
F(\alpha 1_D, 0_{D^c})} = \frac{\mu(A)-\mu(B)}{\mu(C)-\mu(D)}, \quad\forall \alpha>0
\]
\item idempotence (I):
\[
F(\alpha,\ldots,\alpha) = \alpha, \quad\forall \alpha\geq 0,
\]
with restricted versions (I0) for $\alpha=0$ and (I1) for $\alpha=1$. 
\item monotonicity (M), which is non decreasingness of $F$ for each place. 
\end{itemize}
As already noted, (M) implies (M1). All these requirements come from 
considerations linked with the preference of the DM and scales of
measurement. It is possible to show that they are equivalent to a much simpler
set of axioms about $F$.
\begin{prop}
Let $F:\mathbb{R}^n\Rightarrow \mathbb{R}$ and $\mu$ a capacity on $N$. Then
the set of axioms (A1), (A2), (I), (M) is equivalent to the following set of
axioms:
\begin{enumerate}
\item homogeneous extension (HE):
\[
F(\alpha 1_A,0_{A^c}) = \alpha \mu(A), \quad\forall \alpha\geq 0,\forall A\subset N
\]
\item restricted affinity (A) 
\[
F(a_i,0_{\{i\}^c}) = a_i F(1_i,0_{\{i\}^c}), \quad\forall a_i\in \mathbb{R}, \forall i=1,\ldots,n
\]
\item monotonicity (M).
\end{enumerate}
\end{prop}
\textbf{Proof:} $(\Rightarrow)$ Letting $B=D=\emptyset, C=N$ in
(A2) and using (I) lead to $F(\alpha 1_A,0) = \alpha\mu(A)$, which is
(HE). Now, using (A1) with $b_i=d_i=0$, $c_i=1$, $\alpha=1$ and using (I0) we get
$F(a_i,0_{\{i\}^c}) = a_iF(1_i,0_{\{i\}^c})$, which is (A).

$(\Leftarrow)$ Using (A), we get:
\begin{align*}
\frac{F(\alpha a_i,0_{\{i\}^c}) - F(\alpha b_i,0_{\{i\}^c})}{F(\alpha
c_i,0_{\{i\}^c})-F(\alpha d_i,0_{\{i\}^c})} & = \frac{\alpha a_iF(1_i,0_{\{i\}^c}) -
\alpha b_i F(1_i,0_{\{i\}^c})}{\alpha c_iF(1_i,0_{\{i\}^c}) - \alpha d_iF(1_i,0_{\{i\}^c})} \\
& = \frac{a_i - b_i}{c_i - d_i},
\end{align*}
which proves (A1). Now, from (HE)  we get immediately 
\begin{align*}
\frac{F(\alpha 1_A,0_{A^c}) - F(\alpha 1_B, 0_{B^c})}{F(\alpha 1_C,0_{C^c}) -
F(\alpha 1_D, 0_{D^c})} = \frac{\mu(A)-\mu(B)}{\mu(C)-\mu(D)}
\end{align*}
which is (A2). Finally, from (HE) with $A=N$, we get (I) since
$\mu(N)=1$.  $\Box$  

\medskip

Nota: (M) can be dropped from the 2 sets of axioms without changing the
equivalence.

\section{The unicity issue}
Having this simpler set of axioms, we address the question of the unicity of
the solution, i.e. is the \Sipos\ integral the only aggregation function
satisfying the requirements?

First we examine the following extension on $[0,1]^n$ of pseudo-Boolean
functions:
\begin{equation}
\label{eq:pspr}
F(a_1,\ldots,a_n) = \sum_{A\subset N} m(A)\cdot(\bdot_{i\in A}a_i), \forall
a_i\in [0,1]
\end{equation}
as suggested in section \ref{sec:exps}, where $\bdot$ is a
``pseudo-product''. Recall that $m$ is the M\"obius transform of the underlying
capacity. Let us suppose as a basic requirement that $\bdot$ is a commutative
and associative operator, otherwise our expression of $F$ would be ill-defined
since $\bdot_{i\in A} a_i $ would depend on the order of elements in $A$
(commutativity), and on the grouping of elements (associativity). Thus, it is
sufficient to define $\bdot$ on $[0,1]^2$. The following
can be shown.
\begin{prop}
\label{prop:5}
Let $\bdot:[0,1]^2\longrightarrow [0,1]$ be a commutative and associative
operator, and $F$ be given by (\ref{eq:pspr}). Then:
\begin{itemize}
\item[(i)] $F$ satisfies (HE) on $[0,1]^n$ if and only if $\bdot$ coincide with
the product on $\{0,1\}$, satisfies $\alpha\bdot\alpha=\alpha$ for all
$\alpha\in [0,1]$, and $\alpha\bdot 0 = 0$.
\item[(ii)] $F$ satisfies (M) implies $\bdot$ is non decreasing. 
\end{itemize}
\end{prop} 
\textbf{Proof:} (i) ($\Rightarrow$) Let us consider the particular capacity
$u_{1,2}$ defined by $u_{1,2}(A) = 1$ if $\{1,2\}\subset A$, and 0 otherwise
(unanimity game). It is easy to see that its M\"obius transform is such that
$m(\{1,2\})=1$ and 0 elsewhere. Let us consider (HE) with $A=\emptyset$,
$\alpha=1$, and the capacity $u_{1,2}$. We obtain
\[
F(0,\ldots,0) = 1\cdot(0\bdot 0) = u_{1,2}(\emptyset) = 0,
\]
hence $0\bdot 0=0$. Taking now $A=N$, we get:
\[
F(1,\ldots,1) =  1\cdot(1\bdot 1) =  u_{1,2}(N) = 1,
\]
hence $1\bdot 1=1$. Now let us  take $A=\{1\}$, with any $\alpha>0$ and we
obtain from (HE):
\[
F(\alpha,0,\ldots,0) =  1\cdot(\alpha\bdot 0) =  \alpha u_{1,2}(\{1\}) = 0,
\]
hence $\alpha\bdot 0=0$ for any $=\alpha>0$, in particular when $\alpha=1$.
Thus, $\bdot$ coincides with the product on $\{0,1\}$. Lastly, let us apply
(HE) with $A=N$ and again the capacity $u_{1,2}$. We obtain:
\[
F(\alpha,\alpha,\ldots,\alpha) = 1\cdot(\alpha\bdot\alpha) = \alpha
\]  
hence $\alpha\bdot\alpha = \alpha$. 

($\Leftarrow$) For any capacity $\mu$, any $A\subset N$, any $\alpha\in [0,1]$:
\begin{align*}
F(\alpha 1_A,0_{A^c}) & =\sum_{B\subset A}m(B)\cdot(\bdot_{i\in B} \alpha) + 
			\sum_{B\not\subset A}m(B)\cdot[(\bdot_{i\in
			A}\alpha)\bdot(\bdot_{i\not\in A}0)]\\
		& = \alpha\sum_{B\subset A}m(B) + 0\\
		& = \alpha\mu(A).
\end{align*}

(ii) If $\bdot$ is decreasing in some place, and $m$
is positive, then $F$ cannot be increasing, a contradiction. Thus, $\bdot$ is
non decreasing in each place.
$\Box$

\medskip
To go further in the analysis, let us assume in the
sequel that $\bdot$ is  non decreasing. Then we obtain the following result.
\begin{cor}
Let $\bdot:[0,1]^2\longrightarrow [0,1]$ be a commutative, associative, and non
decreasing operator, and $F$ be given by (\ref{eq:pspr}). The following
propositions are equivalent:
\begin{itemize}
\item[(i)] $F$ satisfies (HE), (M) and (A) on $[0,1]^n$.
\item[(ii)] $\bdot$ coincide with
the product on $\{0,1\}$, and satisfies $\alpha\bdot\alpha=\alpha$ for all
$\alpha\in [0,1]$.
\end{itemize}
\end{cor}
\textbf{Proof:} clear from Prop. \ref{prop:5}, the fact that (A) is implied by
(HE)  when working on positive numbers, and the fact that $\alpha\bdot 0=0$ is
implied by $0\bdot 0=0=1\bdot 0$ and non decreasingness. $\Box$

This result gives necessary and sufficient conditions for $\bdot$ in order to
be consistent with our construction. 

Adding the requirement $1\bdot \alpha=\alpha$ for all $\alpha\in[0,1]$,
operator $\boxdot$ becomes a t-norm, as defined in Section
\ref{sec:exps}. Then, the only solution to this set of requirements is the
minimum operator \cite{klmepa00}. Indeed, taking $\alpha,\beta\in [0,1]$ such
that $\alpha\leq \beta$, we have $\alpha=\alpha\bdot \alpha\leq \beta\bdot
\alpha\leq 1\bdot \alpha=\alpha$. This means that the \Sipos\ integral (for
numbers in $ [0,1]$, hence it is the Choquet integral) is the only solution
with this form of pseudo-Boolean function. However, without this additional
assumption, other solutions may exist.

Interestingly enough, the requirement $1\bdot \alpha=\alpha$ has a clear
interpretation in terms of $F$. Indeed, for any $A\subset N$, and any
$\alpha\in [0,1]$,
\begin{align*}
F(1_A,\alpha_{A^c}) & = \sum_{B\subset A}m(B).1 + \sum_{B\not\subset A}m(B).\alpha\\
	& = \sum_{B\subset A}m(B) + \alpha(1- \sum_{B\subset A}m(B))\\
	& = \alpha + (1-\alpha)\mu(A)\\
	& = \alpha+ F((1-\alpha)1_A,0_{A^c}).
\end{align*}
This last expression shows an additivity property of $F$ with particular
acts, specifically:
\[
F(1_A,\alpha_{A^c}) = F((1-\alpha)1_A, 0_{A^c}) + F(\alpha,\ldots,\alpha).
\]
It also shows that $F$ induces a difference scale for those acts, since the
zero can be shifted and set to $\alpha$ without any change.  

\medskip

We now present a solution in the spirit of equation (\ref{eq:sug}), which is in
fact the Sugeno integral \cite{sug74} (see \cite{gra97a}). Let us first
restrict to positive numbers. We introduce the following aggregation function
on $\mathbb{R}^+$:
\begin{equation}
\label{eq:nsug}
S_{m_\vee}(a_1,\ldots,a_n) = \bigvee_{B\subset N}
\Big[m_\vee(B)\cdot \bigwedge_{i\in B} a_i\Big].
\end{equation}
This is a variant of Sugeno integral where the product
takes place of the minimum operator, which satisfies all requirements when
restricted to $\mathbb{R}^+$:
\begin{itemize}
\item monotonicity (M): clear since $m_\vee$ is a non negative set function.
\item (HE): using equation (\ref{eq:mobord}) we get:
\[
S_{m_\vee}(\alpha 1_A,0_{A^c}) = \bigvee_{B\subset A} m_\vee(B)\cdot\alpha = \alpha\cdot\mu(A) = \alpha S_{m_\vee}(1_A,0_{A^c}).
\]
\item (A) for positive numbers is simply a particular case of (HE). 
\end{itemize}  
Note that (HE) works thanks to the product operator in $S_{m_\vee}$. Thus the
original Sugeno integral would not work.

We have to extend this definition for negative numbers in a way similar to the
\Sipos\ integral. The problem of extending the Sugeno integral on negative
numbers has been studied by Grabisch \cite{gra00}, in an ordinal framework. We
adapt this approach to our case and propose the following:
\begin{equation}
\label{eq:ssug}
S_{m_\vee}(a_1,\ldots,a_n) = S_{m_\vee}(a^+_1,\ldots,a^+_n)\svee(-S_{m_\vee}(a^-_1,\ldots,a^-_n))
\end{equation}
with usual notations, and
$\svee$ (called \emph{symmetric maximum}) is defined by:
\[
a\svee b = \left\{	\begin{array}{ll}
			a, & \text{ if } |a|>|b|\\
			0, & \text{ if } b=-a\\
			b, & \text{ otherwise}.
			\end{array}	\right.
\]
The main properties of the symmetric maximum are $a\svee 0 = a$ for all $a\in
\mathbb{R}$ (existence of a unique neutral element), and $a\svee (-a) = 0$ for
all $a\in \mathbb{R}$ (existence of a unique symmetric element). Also, it is
non decreasing in each place, and associative on $\mathbb{R}^+$ and
$\mathbb{R}^-$.

It suffices to verify that (M) and (A) still hold.  (M) comes from
non decreasingness of $\svee$ and $S_{m_\vee}$ for positive arguments. Let us consider $a_i<0$. Then
\[
S_{m_\vee}(a_i,0_{\{i\}^c}) = 0\svee (-a^-_iS_{m_\vee}(1_i,0_{\{i\}^c})) =
a_iS_{m_\vee}(1_i,0_{\{i\}^c}).
\]
Thus the proposed $S_{m_\vee}$ satisfies all requirements of our construction. 

\medskip

Let us examine now a third way to find other solutions. It was suggested in
Section \ref{sec:exps}, formula (\ref{eq:lovss}), which we reproduce here with
suitable notations:
\[
F(a_1,\ldots,a_n) = \sum_{A\subset N} m_1(A)\cdot \bigwedge_{i\in A}
a_i^+ -  \sum_{A\subset N} m_2(A)\cdot\bigwedge_{i\in A} a_i^-, \ \ \
\forall a\in\mathbb{R}^n.
\]  
with $a^+_i:=a_i\vee 0$ and $a^-_i=-a_i\vee 0$. This aggregation function is
built from two different capacities $\mu_1,\mu_2$, one for positive numbers,
and the other one for negative numbers. On each part, it is a Choquet
integral. Let us mention here that this type of
function is well-known in Cumulative Prospect Theory \cite{tvka92}. Obviously,
$F$ satisfies (M) and (HE), let us check (A) for negative numbers. We have for
any $i\in N$, any $a_i<0$:
\[
F(a_i,0_{\{i\}^c}) = 0 - m_2(\{i\})a^-_i = a_im_2(\{i\}).
\]
But $F(1_i,0_{\{i\}^c}) = m_1(\{i\})$, so that a necessary and
sufficient condition to ensure the compatibility with our construction is:
\[
m_2(\{i\}) = m_1(\{i\}), \quad\forall i\in N.
\]

At this stage, we do not know if other solutions exist, and a complete
characterization is left for further study.

\section{Conclusion}
We have shown in this paper that considering, besides classical comparative
information, absolute information, strongly modifies the aggregation problem in
MCDA. The classical multilinear model is no more adequate but new models like
Choquet and \Sipos\ integrals appear because absolute information allows to
lead to commensurable scales. Among these two models, we have shown that the
\Sipos\ integral is the only acceptable solution, although there exist other
models fitting all the requirements. The approach leading to the unicity of the
solution based on \Sipos\ integral is deserved for a subsequent study.

 {\small
\bibliographystyle{plain}

\bibliography{../BIB/fuzzy,../BIB/grabisch,../BIB/general}

\end{document}